\documentclass[usenatbib,useAMS]{mn2e}
\usepackage{times}
\usepackage{amsmath}	
\usepackage{amssymb}

\usepackage{blindtext}
\usepackage{graphicx}
  \let\url\relax

\def\apjs{{ApJS}} 
\def\apjl{{ApJL}}
\def\aap{{\em A.\&A}}

\newcommand{\be}{\begin{equation}}
\newcommand{\ba}{\begin{eqnarray}}
\newcommand{\ee}{\end{equation}}
\newcommand{\ea}{\end{eqnarray}}  

\def\lesssim{\mathrel{\hbox{\rlap{\hbox{\lower4pt\hbox{$\sim$}}}\hbox{$<$}}}}
\def\gtrsim{\mathrel{\hbox{\rlap{\hbox{\lower4pt\hbox{$\sim$}}}\hbox{$>$}}}}

\def\gtsima{$\; \buildrel > \over \sim \;$}
\def\ltsima{$\; \buildrel < \over \sim \;$}
\def\gsim{\lower.5ex\hbox{\gtsima}}
\def\lsim{\lower.5ex\hbox{\ltsima}}
\def\simgt{\lower.5ex\hbox{\gtsima}}
\def\simlt{\lower.5ex\hbox{\ltsima}}
\def\simpr{\lower.5ex\hbox{\prosima}}
\def\la{\lsim}
\def\ga{\gsim}

 \def\msun{{M_\odot}}

\def\simless{\mathbin{\lower 3pt\hbox
   {$\rlap{\raise 5pt\hbox{$\char'074$}}\mathchar''7218$}}}   
\def\simgreat{\mathbin{\lower 3pt\hbox
   {$\rlap{\raise 5pt\hbox{$\char'076$}}\mathchar''7218$}}}   

\newcommand{\myfrac}[3][0pt]{\genfrac{}{}{}{}{\raisebox{#1}{$#2$}}{\raisebox{-#1}{$#3$}}}

\title[Stochastic bias] {Modelling the stochasticity of
  high-redshift halo bias}
\author[Nasirudin et al.]
         {Ainulnabilah Nasirudin$^{1,2,3}$, Ilian~T.~Iliev$^3$, Kyungjin Ahn$^4$\\
           $^1$ International Centre for Radio Astronomy Research - Curtin University, 1 Turner Avenue, Bentley WA 6102, Australia \\
$^{2}$ARC Centre of Excellence for All Sky Astrophysics in 3 Dimensions (ASTRO 3D)\\         
$^3$Department of Physics \& Astronomy, University of Sussex,
  Brighton, BN1 9QH, UK\\
  $^4$Department of Earth Sciences, Chosun University, Gwangju, Korea}
 
\begin{document} \date{Draft - \today}

\pubyear{2019}\volume{0000}

\maketitle

\label{firstpage}

\begin{abstract} 
A very large dynamic range with simultaneous capture of both 
large- and small-scales in the simulations of 
cosmic structures is required for correct modelling of many
cosmological phenomena, particularly at high redshift. This is
not always available, or when it is, it makes such simulations
very expensive. We present a novel sub-grid method for modelling
low-mass ($10^5\,M_\odot\leq M_{\rm halo}\leq 10^9\,M_\odot$) haloes, which
are otherwise unresolved in large-volume cosmological simulations
limited in numerical resolution. In addition to the deterministic
halo bias that captures the average property, we model its
stochasticity that is correlated in time. We find that the instantaneous
binned distribution of the number of haloes is well approximated by a
log-normal distribution, with overall amplitude modulated by this
``temporal correlation bias''. The robustness of our new scheme is tested
against various statistical measures, and we find that temporally
correlated stochasticity generates mock halo data that is significantly
more reliable than that from temporally uncorrelated stochasticity.
Our method can be applied for simulating processes that depend on
both the small- and large-scale structures, especially for those
that are sensitive to the evolution history of structure
formation such as the process of cosmic reionization.
As a sample application, we generate a mock distribution of medium-mass 
($ 10^{8} \leq M/M_{\odot} \leq 10^{9}$) haloes inside a 500 Mpc$\,h^{-1}$,
$300^3$ grid simulation box. This mock halo catalogue bears a reasonable
statistical agreement with a halo catalogue from numerically-resolved
haloes in a smaller box, and therefore will allow a very self-consistent
sets of cosmic reionization simulations in a box large enough to generate
statistically reliable data.
\end{abstract}

\begin{keywords}
  haloes --- cosmology: theory --- dark ages, reionization, first stars --- methods: numerical
\end{keywords}


\section{Introduction}

Cosmological haloes are the sites where most active astrophysical
processes occur. Understanding the formation, evolution and spatial
distribution of cosmological haloes thus allows theoretical access
to many astrophysical phenomena and links to observational cosmology.
The halo information is typically summarised in the form of a halo
catalogue. Modelling galaxy surveys in observational cosmology and
studies of the process of cosmic reionization, which generates
observational features at cosmological scales, are both examples
that can benefit from reliable, realistic halo catalogues.

For cosmic reionization modelling in particular, properly understanding
 the nature of the ionising sources responsible for the Epoch of
 Reionization (EoR) is an important problem which can aid in interpreting
current and future detection experiments. Minihaloes, small haloes in which gas 
cannot cool through atomic line cooling, are generally conceived 
to be the site of the first, Pop.~III, stars responsible for the 
early stages of EoR \citep[e.g.][]{barkana2001beginning}. At the 
same time, both the stellar-mass binaries and the progenitors of 
super-massive black holes are believed to be among the main 
sources of X-ray radiation, which heats the neutral inter-galactic 
medium (IGM) unreachable by UV radiation from stars 
\citep{ricotti2004x,madau2004early,knevitt2014heating}. The later 
stages of re-ionisation are thought to be dominated by emission 
from more massive galaxies, with halo masses above about $10^8\,M_\odot$, 
which we refer to as atomically cooling haloes (ACHs) based on 
their dominant gas cooling mechanism. These can in turn be split 
into low-mass ACH (LMACHs), with masses less than about 
$10^9\,M_\odot$, for which the gas accretion onto them can be 
suppressed by radiative feedback, and the larger, high-mass
 ACH (HMACHs) which are largely unaffected by such feedback 
\citep{2007MNRAS.376..534I,dixon2016large,ahn2015non}. The process of reionization 
is believed to be inhomogeneous and anisotropic on large scales 
\citep{2014MNRAS.439..725I}, hence a large-volume simulation box is
needed to capture the extent of the processes involved. However, this
also means that cosmic reionization simulations often have a relatively
low resolution, resulting in the need for sub-grid modelling of the
low-mass haloes in which the majority of reionization sources reside.

One perspective towards modelling the sub-grid halo population is the 
peak-background split scheme \citep{1986ApJ...304...15B} which provides
a framework for understanding how haloes form in a way biased toward
high-density environment. In this approach, a halo is associated with the linear
overdensity $\delta_{\rm lin}$ that satisfies the halo formation criterion,  
$\delta_{\rm lin}>\delta_{c}$, where $\delta_{c}\approx 1.636$ is the
critical overdensity. This put theoretical foundations under the
well-known Press-Schechter (PS) formalism \citep{1974ApJ...187..425P},
whose fudge multiplicity factor 2 was later explained rigorously by
\citet{Bond1991} through their extended Press-Schechter formalism (or
the excursion set formalism). The halo bias due to a large-scale density
environment specified by its linear overdensity $\Delta_{\rm linear}$ can
also be naturally accounted for in this framework, because the small-scale
density fluctuation now only needs to satisfy the modified criterion
$\delta_{\rm lin}>\delta_{c}-\Delta_{\rm linear}$ \citep{Cole1989}. 
\citet{Mo1996} calculated a fully nonlinear bias prescription, by
combining this peak-background split scheme with the spherical top-hat
collapse model.

A notable shortcoming  of both the average PS halo mass function and
the nonlinear halo bias model by \citet{Mo1996} is that the number of
haloes predicted this way does not match the numerically simulated
haloes well, especially massive, rare haloes at any given epoch. The
discrepancy in the average halo mass function stimulated a set of
fitting functions based either on a more detailed theory in the halo
formation, supported by simulation data (e.g. \citealt{Sheth1999}, ST
hereafter) or empirical fits to numerical simulations
(\citealt{2001MNRAS.321..372J,Warren2006,Reed2007,Lukic2007,Lim2013,
  Watson2014}). The discrepancy in the biased halo mass function seems
to be resolved by a simple yet attractive solution by \citet[BL
  hereafter]{2004ApJ...609..474B}, which is a hybrid scheme of
combining the ST mass function and the bias prescription from
the extended PS formalism. \citet{ahn2015non} extended this idea to
combine the average mass function of simulated haloes, instead of the
ST mass function, and the bias prescription from the extended PS
formalism, and have found that this scheme has an excellent predictive
power on the nonlinear bias of haloes in the high-redshift regime.

In order to add more naturality to the sub-grid modelling, however,
the ``deterministic'' prescription described so far is not sufficient,
and the stochasticity of halo bias also needs to be implemented. This
stochasticity does not follow the pure Poisson distribution, because
the correlation of haloes at the sub-grid level produces a variance in
the number of haloes, in addition to the usual shot noise
\citep{1993ppc..book.....P,Dekel1999}. In the presence of the sub-grid
correlation, this additive variance extends the tails of the distribution
function of the halo number, or ``super-Poissonian'' distribution, which
is well fitted by functional forms by \citet{Saslaw1984} and
\citet{Sheth1995}.

The sub-grid modelling of haloes is naturally connected to the effort
to generate mock halo catalogues under the knowledge of large-scale
density field. This idea is implemented in PINOCCHIO
\citep{Monaco2002,Monaco2013} and PTHALOES \citep{Scoccimarro2002,
  Manera2013}, which generate mock halo catalogues based on the
quasi-linear density field. With an adaptive high-order perturbation
theory, mock catalogues can be generated even more precisely  (e.g.
Patchy by \citealt{Kitaura2014} in which they used the 2nd-order
Lagrangian perturbation theory but at the same time improved on
mitigating the problems caused by the unwanted free-streaming of
particles in small scales). The feasibility to use such a prescription
to study e.g. the Baryon Acoustic Oscillation (BAO) feature in density
power spectrum has been presented \citep{Kitaura2014}.

We find that one key ingredient is still missing in sub-grid
modelling efforts described above. This is the {\em temporally correlated
stochasticity}: stochasticities in the halo distribution at mutually
nearby epochs should be correlated. This would not be crucial if one
is interested only in a limited range of redshifts. For example, the
study of BAO through galaxy surveys at low redshift may need to
focus only on the fields of halo population that is instantaneous or
mildly changing in time. However, in cases where continuous evolution
is important, correlation of stochasticity in time is also
crucial. For example, in the study of cosmic reionization, how haloes
are generated in time and space are cumulatively imprinted in the late
phase of the process in terms of the morphology of H II regions.
Therefore, in this paper, we present our quantitative study on
the temporally correlated stochasticity and the feasibility to use
this prescription to self-consistently generating mock halo catalogues
both in time and space. Previously we have implemented the deterministic
bias only, without stochasticity, as sub-grid treatment. Nevertheless,
through this method, we have found that small-mass haloes impact the
physics of reionization quite substantially: minihaloes yield extended
and self-regulated reionization epoch (\citealt{2012ApJ...756L..16A};
even found favoured by the CMB polarization data by Planck:
\citealt{Heinrich2017}), the observed 3-dimensional (3D) imaging and
21-cm power spectrum depends on the minimum halo mass
\citep{Dixon:2016aa,2018MNRAS.473.2949G}, and statistics of neutral gas
islands in the late phase of islands will be affected by LMACHs
\citep{2019arXiv190301294G}, to give a few examples. Application of a
self-consistently calculated stochasticity is therefore expected to
provide a more reliable picture on the physics of cosmic reionization.

This paper is organized as follows. We describe our method to generate
halo bias that can assimilate the N-body simulation data in
\S~\ref{sect:method}. We present mock realizations of haloes in those
boxes that resolve haloes of given mass range, and compare the resulting
statistical measures of the actual N-body data and the mock data in
\S~\ref{sect:tests}. An application of our method to a large,
500$\,h^{-1}$ Mpc box with a test on its validity is described in
\S~\ref{sect:eor_application}. We summarise our results and discuss
relevant issues in \S~\ref{sect:summary}.

\section{Methodology}
\label{sect:method}

\subsection{The N-body simulation data}

\begin{figure*}
\centering
\includegraphics[width=\textwidth]{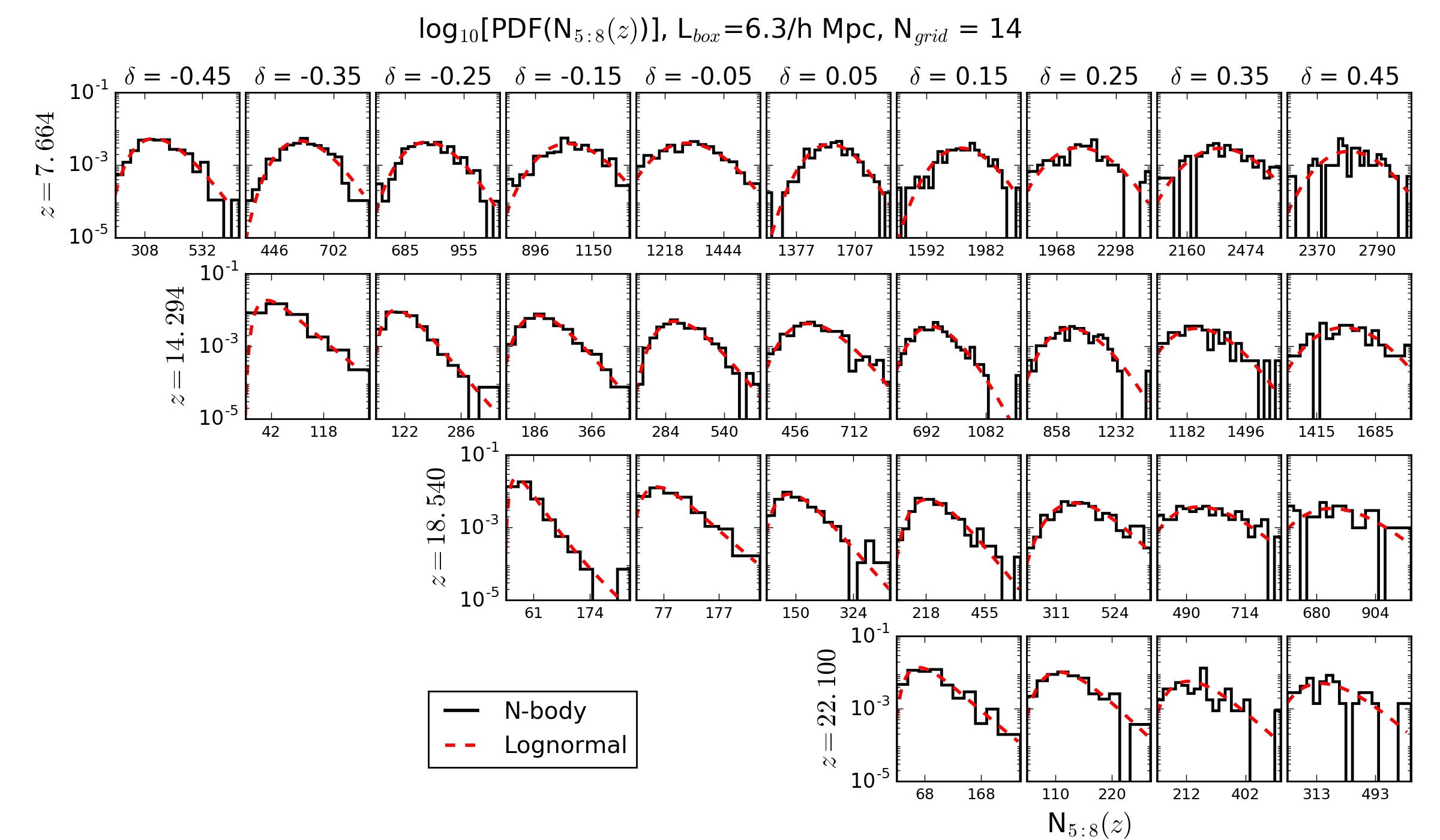}
\caption{\small\label{fig:pdf14} PDF of the number of MHs, $N_{5:8}$,
  contained within cells of given overdensity $\delta$ on our $14^3$-grid
  $6.3\,h^{-1}$~Mpc-box. $\delta$ is shown on top of each subplot while
  redshifts are shown on the left. The PDF of the N-body halo data (black
  solid) is fit by a log-normal distribution (red dashed).} 

\centering
\includegraphics[width=\textwidth]{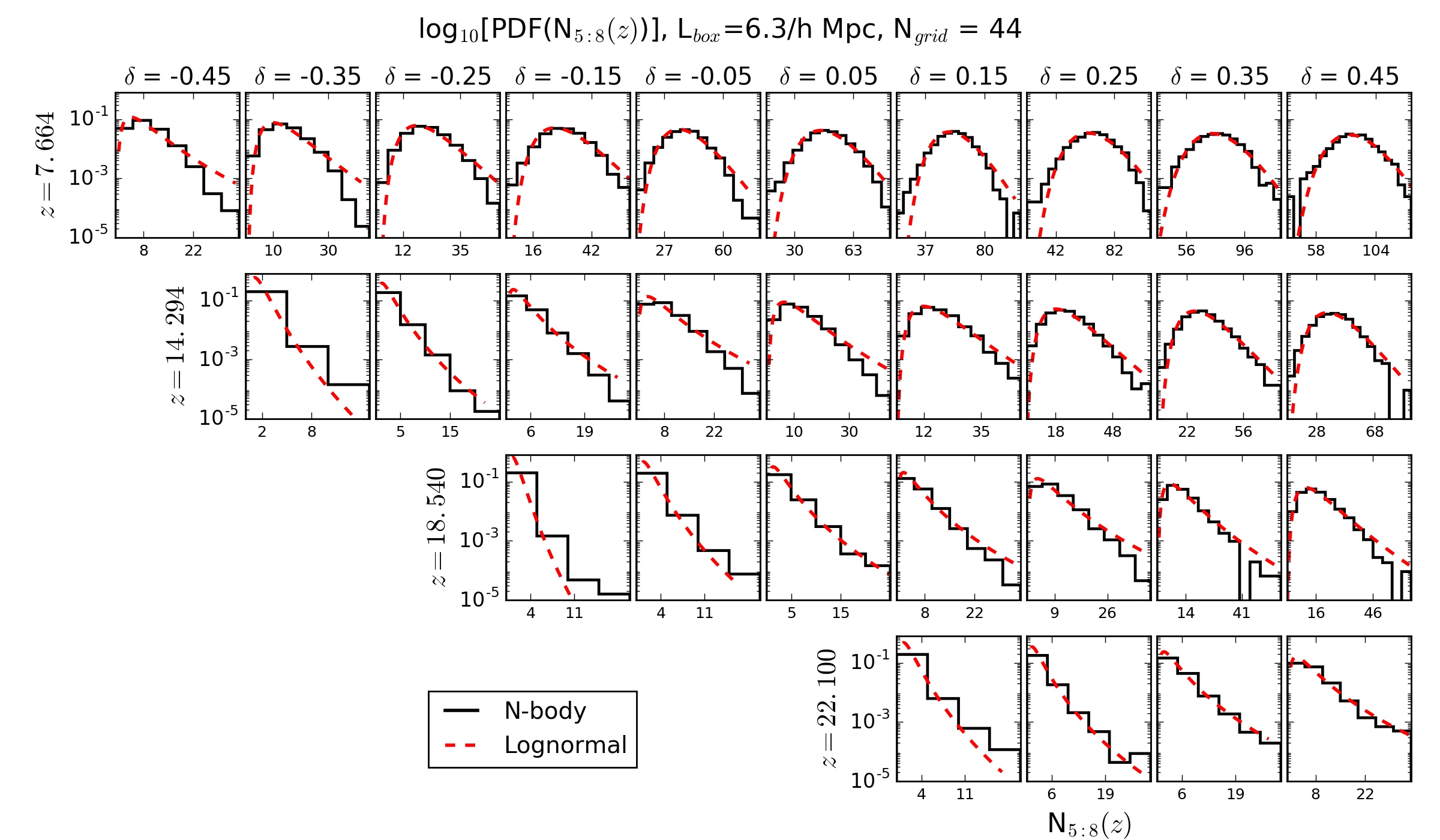}
\caption{\small\label{fig:pdf44} Same as in Fig.~\ref{fig:pdf14}, but with
  the number of haloes sampled on a $44^3$ grid.}
\end{figure*}

\begin{figure*}
\centering
\includegraphics[width=\textwidth]{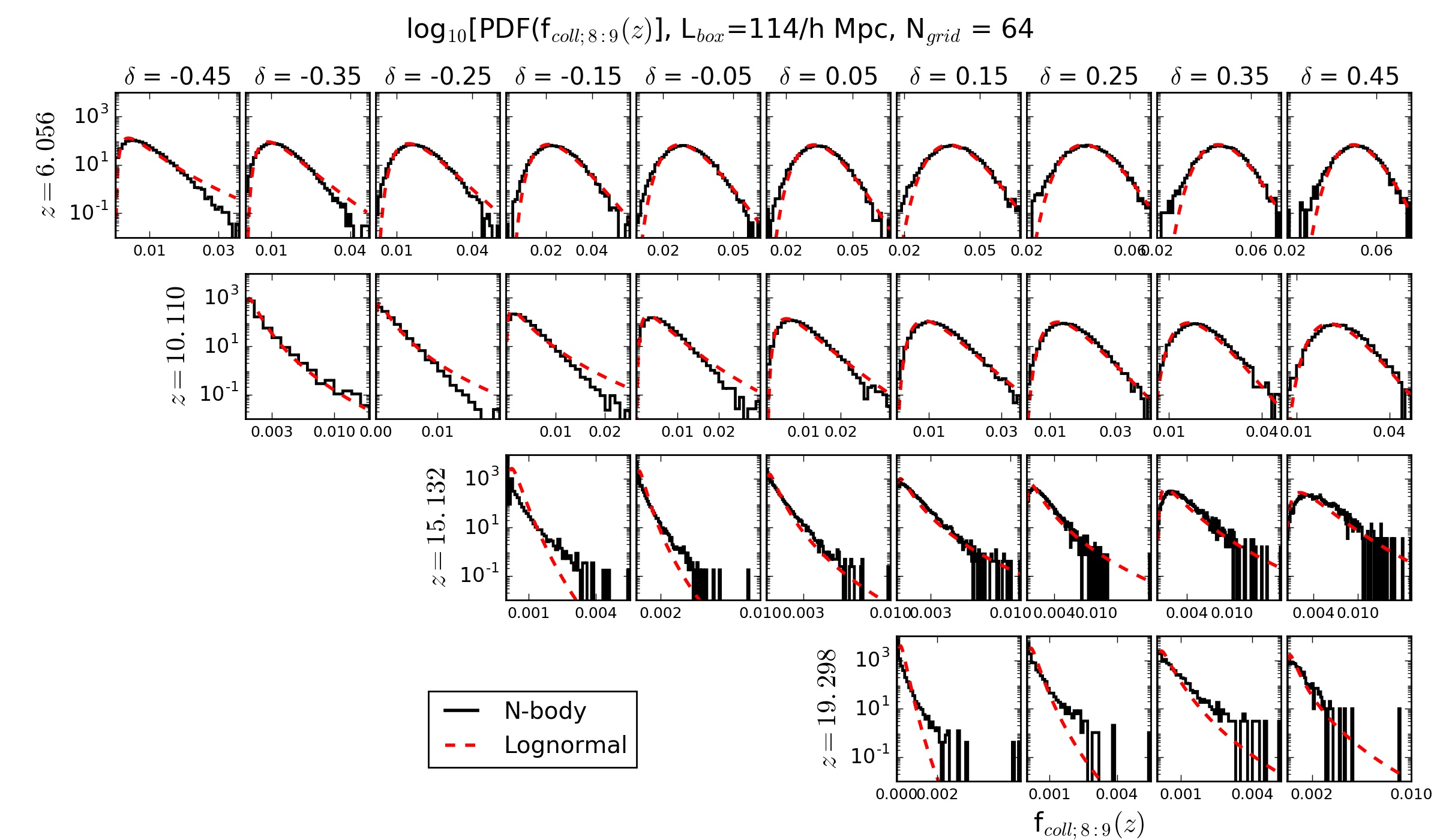}
\caption{\small\label{fig:pdf64} PDF of the LMACH collapsed fraction,
  f$_{\rm coll,8:9}$, contained within cells of given overdensity $\delta$
  on our $64^3$-grid $114\,h^{-1}$~Mpc-box. Plotting convention is the
  same as in Figure~\ref{fig:pdf14}.}

\centering
\includegraphics[width=\textwidth]{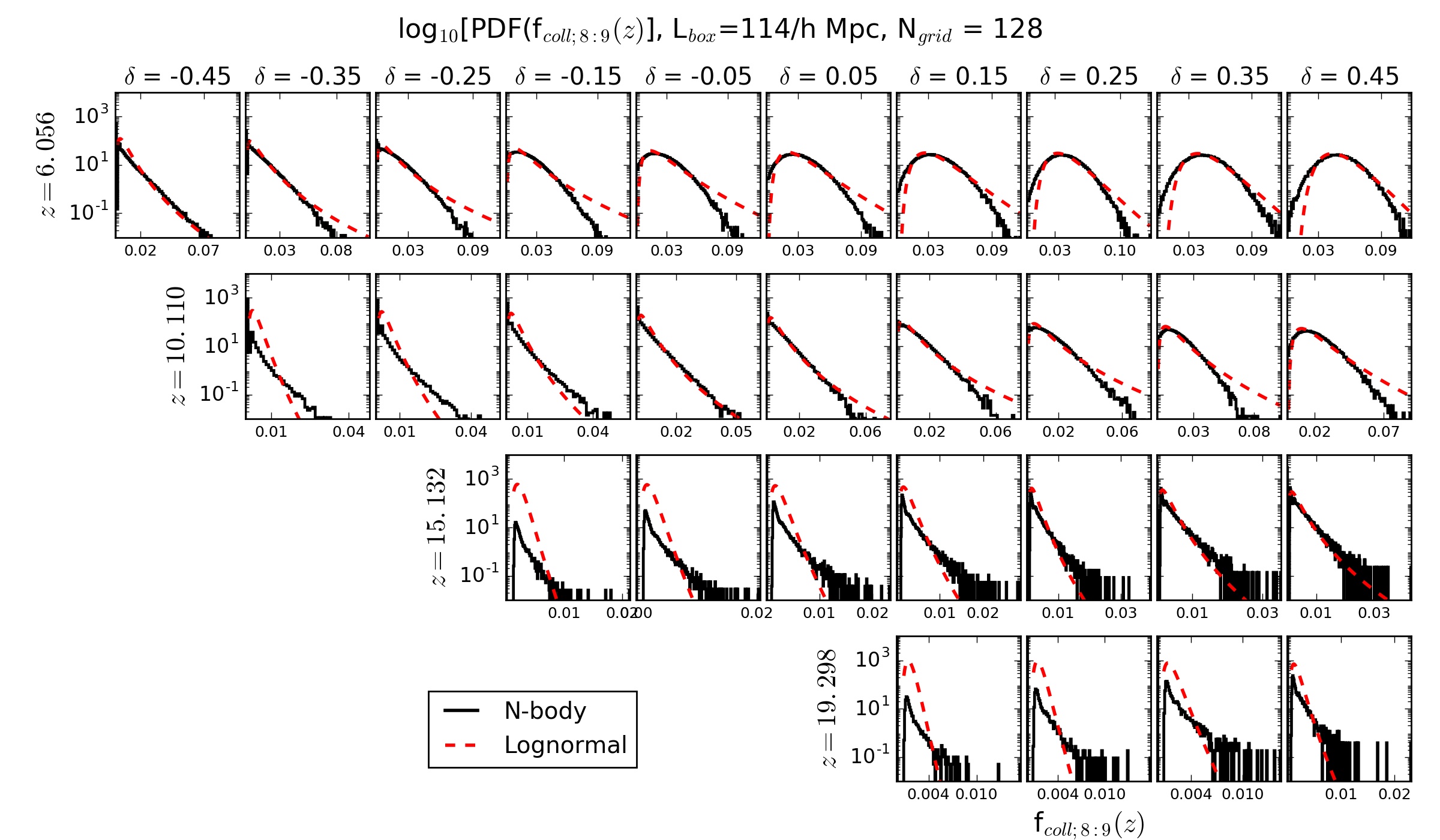}
\caption{\small\label{fig:pdf128} Same as in Fig.~\ref{fig:pdf64}, but
  with the collapsed fraction sampled on a $128^3$ grid.}
\end{figure*}

Halo bias is composed of the deterministic bias and the stochastic
bias. The former represents the average conditional probability that
cells with given overdensity $\delta$ will host haloes of given mass
range. The latter represents the unavoidable stochasticity, due to
dependencies beyond local density, in the number (or collapsed
fraction) of haloes across cells with given $\delta$. These two
components can be expressed in a single probability distribution
function in terms of the average quantity $\mu$ and at the least
the standard deviation $\sigma$ (that of the approximately log-normal
distribution in our case as in Equation~\ref{eqn:lognorm}), respectively. 

Motivated by the physics of the radiation sources during the Cosmic Dawn
and EoR, we split the dark matter haloes into three physically-motivated
mass ranges \citep[see e.g.][]{2007MNRAS.376..534I}: high-mass
atomically-cooling haloes (HMACHs), defined by $M_{\rm halo}>10^9M_\odot$;
low-mass atomically-cooling haloes (LMACHs), with
$10^8M_\odot<M_{\rm halo}<10^9M_\odot$, and minihaloes (MHs), with
$M_{\rm halo}<10^8M_\odot$. We focus on implementing the stochasticity of the
latter two types in large-scale simulations, since HMACHs are relatively
more easily resolved and thus sub-grid modelling is usually unnecessary,
although our model is easily extensible to such haloes as well. By
definition, ACHs form stars efficiently by cooling the gas through atomic
line radiation, and the star formation therefore is expected to be roughly
proportional to the total amount of gas available, which in turn is
proportional to the collapsed fraction in such haloes. In contrast, gas in
MHs is easily disrupted and photo-evaporated by the first star(s) to form
within each halo, or in nearby haloes, thus we expect that each MH forms
just one, or at most a few stars, irrespective to its actual total gas mass.
Therefore, in our modelling for LMACHs we consider the local collapsed
fraction, while for MHs we consider the local number of haloes. Our model
is applicable to either quantity.

The simulation data used throughout this work is based on a suite of
N-body simulations using the CubeP$^3$M code \citep{2013MNRAS.436..540H}.
The MH data is based on a high-resolution simulation in a box of
6.3$\,h^{-1}$~Mpc per side, with $1728^3$ particles, force resolution of
$182\,h^{-1}$~pc and minimum halo mass resolved (with 20 particles or more)
of $\sim10^5\,M_\odot$. This roughly corresponds to the halo 'filtering mass'
\citep{2000ApJ...542..535G} below which haloes struggle to keep their gas
content even before reionization. The LMACHs data is based on a simulation
in a box of 114$\,h^{-1}$~Mpc per side, with $3072^3$ particles, force
resolution of $1.8\,h^{-1}$~kpc and minimum resolved halo mass of
$\sim10^8\,M_\odot$. The density field is smoothed onto a uniform grid of
14$^3$ and 44$^3$ cells when sampling MHs (hereafter $N_{5:8}$), and into
64$^3$ and 128$^3$ cells when sampling the {\em local} collapsed fraction
of LMACHs ($f_{\rm coll,8:9}$), given by
\begin{equation}
f_{\rm coll,8:9}= \myfrac[3pt]{M_{8:9}}{M},
\end{equation}
where $M_{8:9}$ is the mass of haloes between $10^8$ and $10^9$M$_\odot$ and
$M$ is the mass of a cell. The local overdensity, $\delta$, of each cell is
defined by
\begin{equation}\label{eqn:delta}
\delta=\frac{\rho}{\bar{\rho}}- 1,
\end{equation}
where $\rho$ is the average density of a  cell and $\bar{\rho}$ is the mean
density of the universe.

\subsection{Instantaneous halo bias in N-body simulation}
\label{sect:instant_Nbody}

We empirically quantify the instantaneous halo bias ("instantaneous bias"
hereafter) from N-body simulation data. For now we ignore any possible
temporal correlation but just consider the local cell overdensity $\delta$
and redshift $z$. First, we sample $\delta$'s of cells with the bin width
$\Delta \delta=0.1$. This guarantees a reasonable amount of sampling of grid
cells for any $\delta < 10$. Only for the highest-$\delta$ cells which are
rare, we enlarge the bin size substantially: cells of  $\delta>10$ are
grouped into two coarse bins -- $10 \leq \delta \leq 15$ and $\delta > 15$.
Second, for each $\delta$ bin, we measure the empirical Probability Density Function (PDF) of $N_{5:8}$ or
$f_{\rm coll,8:9}$ by visiting all cells of the given $\delta$.\\

Any conditional parameters hereafter
denote parameters measured in appropriate bins; e.g. $\mu(z|\delta=0.5)$ is
the average value of all $\mu(z|\delta)$'s when
$\delta \in [0.5-\Delta \delta/2, \,0.5+\Delta \delta/2]=[0.45,\,0.55]$. . The following terminology and parameters are useful in describing the
empirical, instantaneous stochasticity:
\begin{itemize}
\item $x$: the value of either $N_{5:8}$ or f$_{{\rm coll},8:9}$.
\item $x_{\rm min}(z|\delta)$, $x_{\rm max}(z|\delta)$: the minimum and maximum,
  respectively, of $x$ found in cells of overdensity $\delta$ at redshift $z$.
\item $\mu(z|\delta)$: the average of $\ln(x)$, given cells of $\delta$ at $z$.
\item $\sigma(z|\delta)$: the standard deviation of $\ln(x)$, given cells of
  $\delta$ at $z$.
\item $N_{+}(z)$: the total number of non-empty cells at $z$.
\item $N_{0}(z)$: the total number of empty cells at $z$.
\item $N_{+}(z|\delta) $: the number of cells of $\delta$ which are non-empty
  at redshift $z$.
\item $N_{0}(z|\delta)$: the number of cells of $\delta$ which are empty at
  $z$.
\end{itemize}

The empirical results for PDFs of $N_{5:8}$ are shown in Figures \ref{fig:pdf14}
and \ref{fig:pdf44}, and those for PDFs of $f_{\rm coll,8:9}$ are shown in
Figures~\ref{fig:pdf64} and \ref{fig:pdf128}. The PDFs are roughly Gaussian
close to its peak, but have considerable skewness away from it, and are thus
overall not well represented by a Gaussian. This ''super-Poissonian''
distribution is caused by the non-zero auto-correlation of halo population
in the sub-cell scale \citep{Peebles:1980aa,ahn2015non}, which can be well-fit
by distribution functions suggested by \citet{Saslaw1984} and
\citet{Sheth1995}. Instead, we employ a lognormal distribution, with which
the skewness is easily realized with just two parameters -- the average and
the standard deviation of the logarithmic -- to fit the empirical data:
\begin{equation}
  \label{eqn:lognorm}
f(x, z |\delta) = \frac{1}{x} \frac {1}{\sigma {\sqrt {2\pi }}}\exp^{-{\frac {\left(\ln x-\mu \right)^{2}}{2\sigma ^{2}}}},
\end{equation}
where $\mu=\mu(z|\delta)$ and $\sigma=\sigma(z|\delta)$. We use this
lognormal PDF to represent the PDF of non-empty cells only, since empty
cells are not of interest here and including them would distort distribution.
Note that we do not take ($\mu$, $\sigma$) as free parameters to find the
best fit to the empirical data, but instead use the empirical values of
($\mu(z|\delta)$, $\sigma(z|\delta)$) and consider the goodness of the
fit later.

The lognormal fits (red dashed lines in Figs.~\ref{fig:pdf14} - \ref{fig:pdf128}) largely
match the empirical distributions well. In both cases of MHs and LMACHs,
the lognormal fit works better for the larger cell size: $N_{\rm grid}=14$
case (Fig.~\ref{fig:pdf14}) is better than $N_{\rm grid}=44$ case
(Fig.~\ref{fig:pdf44}) for MHs, and $N_{\rm grid}=64$ case (Fig.~\ref{fig:pdf64})
is better than $N_{\rm grid}=128$ case (Fig.~\ref{fig:pdf128}) for LMACHs. The
mismatch occurs in many cases, but only in the tails of PDFs where the
fractional contribution becomes relatively unimportant. Only at high-redshift
($z\gtrsim 15$), small cell-size cases for LMACHs show the biggest mismatch
(Fig.~\ref{fig:pdf128}), even in its amplitude. Overall, we expect that any
resulting statistical measures from this lognormal fitting will be only
slightly different from those of the N-body data. 

\subsection{Implementing instantaneous halo bias}
\label{sect:instant_mock}

We now describe our scheme to realize instantaneous bias for generating
mock halo catalogues using the empirical parameters, described in section
\ref{sect:instant_Nbody}, as the basis. A fluctuating 3D density field
should be provided at a target redshift, ideally by numerical simulations
that resolve the nonlinear density environment for given filtering scale
(Eulerian cell size).

Our algorithm is described by the sequence below. Any quantity with a prime
symbol denotes a value related to the mock data.
\begin{enumerate}
\item Once the N-body particle density field is interpolated onto a uniform
  grid, group the grid cells according to discrete bins of $\delta$. 
\item Among the cells of given $\delta$, randomly choose a fraction
  $P_{+}(z|\delta)$ of these cells that will host haloes, with the
  ``conditional occupation probability''
\begin{equation}\label{eqn:P+_instant}
P_{+}(z|\delta) = \frac{N_{+}(z|\delta) }{N_{+}(z|\delta) + N_{0}(z|\delta)},
\end{equation}
and leave the remaining cells devoid of any haloes. 
The number of non-empty cells of $\delta$ in the given density field found
this way, $N'_{+}(z|\delta)$, may differ from $N_{+}(z|\delta)$ in general,
because $N'_{+}(z|\delta)=P_{+}(z|\delta)\,N'(z|\delta)$ where $N'(z|\delta)$
is the number of grid cells of $\delta$ in the given density field. The total of
 $N'_{+}(z|\delta)$, ($N'_{+}(z)$) however, should follow
 \begin{equation}\label{eqn:new_tot}
N'_{+}(z)=\frac{N_{+}(z)}{N_{\textrm{grid}}}\times N'_{\textrm{grid}}.
\end{equation}
\item Use Monte Carlo sampling of $x$ based on Equation (\ref{eqn:lognorm})
  to populate these non-empty cells with haloes. Sample $x$ from the bounded
  range $x=[ x_{\rm min}(z|\delta), x_{\rm max}(z|\delta)]$.
\item When $N'_{+}(z|\delta)\lesssim 10$ it is inappropriate to use the Monte
  Carlo sampling due to rarity of such cells; in such cases the empirical data
  shows convergence of $x$ to $\mu(z|\delta)$. If this happens, only use the
  deterministic bias, by setting $x= e^{\mu(z|\delta)}$ in those cells.
\end{enumerate}

This scheme only considers the instantaneous information, and thus can be
easily applied once the empirical parameters described in
\S~\ref{sect:instant_Nbody} are available. Note that when the range
of $\delta$ in the given density field exceeds the range of the empirical
values due to e.g. the increased size of a simulation box, one needs to
extrapolate the empirical parameters for those outliers of $\delta$. 

\subsection{Temporal halo bias in N-body simulation}
\label{sect:temporal_Nbody}

Just as in the case of the instantaneous bias, we first find an empirical
model for the temporal stochastic halo bias ("temporal bias" hereafter)
from N-body halo data, and then develop a method to generate mock halo
catalogues based on this model. Ideally, it would be best to find e.g. a
universal relation between the redshift $z$ and the degree of stochasticity
for any given Eulerian or Lagrangian cell. Unfortunately we could not find
such a clean relation yet, but instead found a model that reflects temporal
stochasticity of N-body halo catalogues to a significant extent. We will
show how well this method mimics the actual N-body halo catalogues in
\S~\ref{sect:results} through various statistical measures.

Our empirical model for temporal stochasticity can be described by the
following parameters, where appropriate binning is assumed for conditional
values:
\begin{itemize}
\item $\Delta z_{+,i}$: the redshift interval during which an Eulerian
  cell $i$ is not empty.
\item $\Delta z$: the full duration of our N-body simulation, under the
  assumption that haloes of interest start to emerge in the simulation
  volume from the starting redshift.
\item $\bar{\delta}_{i}$: the average overdensity of the cell $i$ over
  $\Delta z$.
\item $f_{+,i}\equiv \Delta z_{+,i}/\Delta z$: the fraction of time during
  which the cell $i$ is not empty. 
\item $f_{+}(\bar{\delta})$: the time fraction that an Eulerian cell with
  mean overdensity $\bar{\delta}$ is not empty. Here, $\bar{\delta}$ is
  the average of $\bar{\delta}_{i}$ of cells having $f_{+,i}=f_{+}$, and thus
  this function relates $f_{+}$ and $\bar{\delta}$ in averaged, deterministic
  way.
\item $N_{+}(z|x_{\rm prev}, +)$: the total number of non-empty cells at $z$,
  among those cells which had $x=x_{\rm prev}\equiv x(z_{\rm prev}) > 0$. Here,
  $z_{\rm prev}$ is the redshift of the N-body data recorded just before the
  redshift $z$.
\item $N_{0}(z|x_{\rm prev}, +)$: the total number of empty cells at $z$, among
  those cells which had $x=x_{\rm prev} > 0$.
\end{itemize}

\begin{figure*}
\centering
\includegraphics[width=0.84\textwidth]{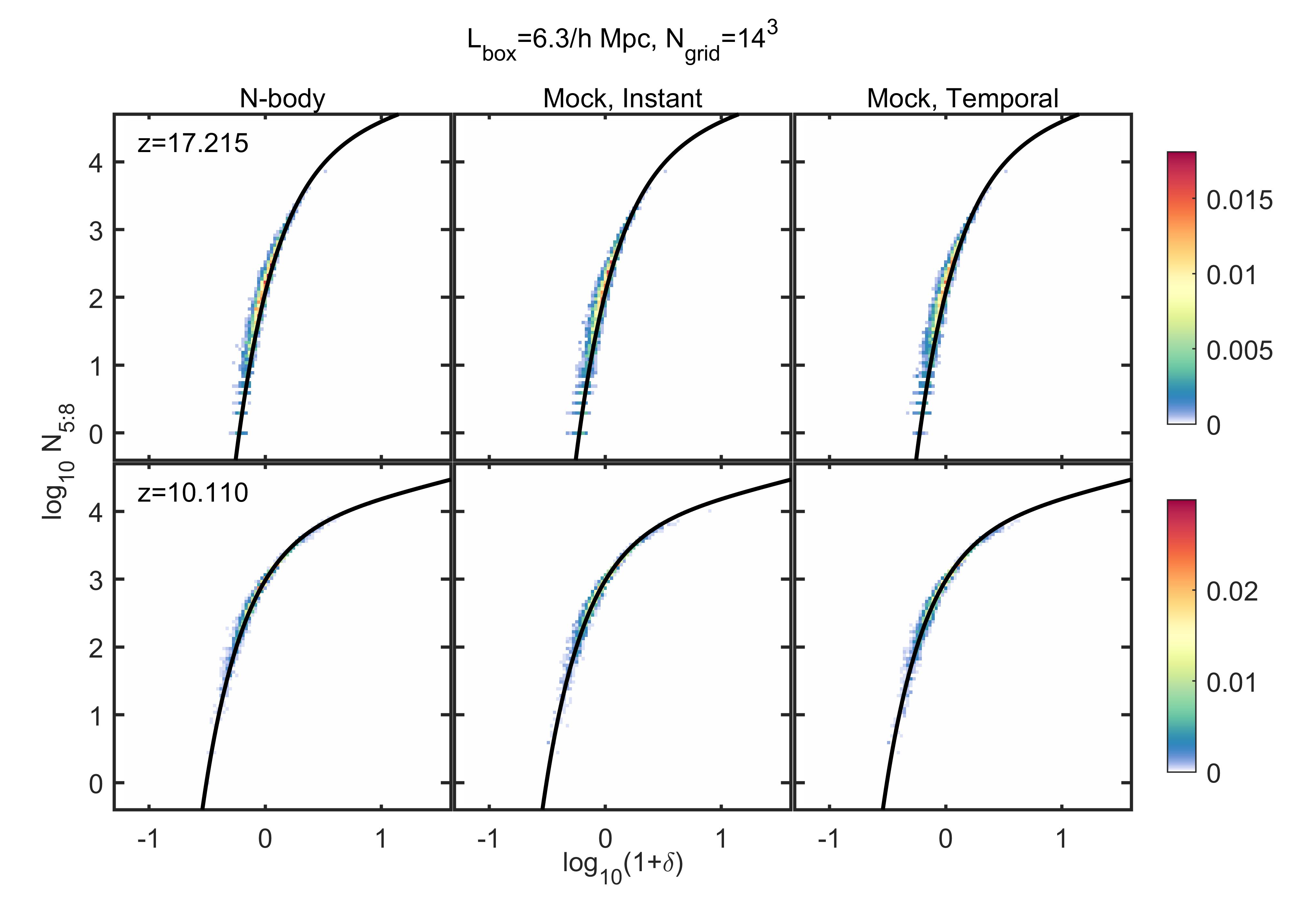}
\caption{\small\label{fig:num14} Number of MHs, $N_{5:8}$, per cell vs cell
  overdensity $\delta$ at $z=17.215$ (upper panel) and $z=10.110$ (lower
  panel) for 6.3 Mpc/h box and 14$^3$ grid (cell size $450\,\rm h^{-1}kpc$).
  Shown are the N-body halo data (left panels), the instantaneous mock halo (middle panels), and temporal mock halo (right panels).
}
\end{figure*}

\begin{figure*}
\centering
\includegraphics[width=0.84\textwidth]{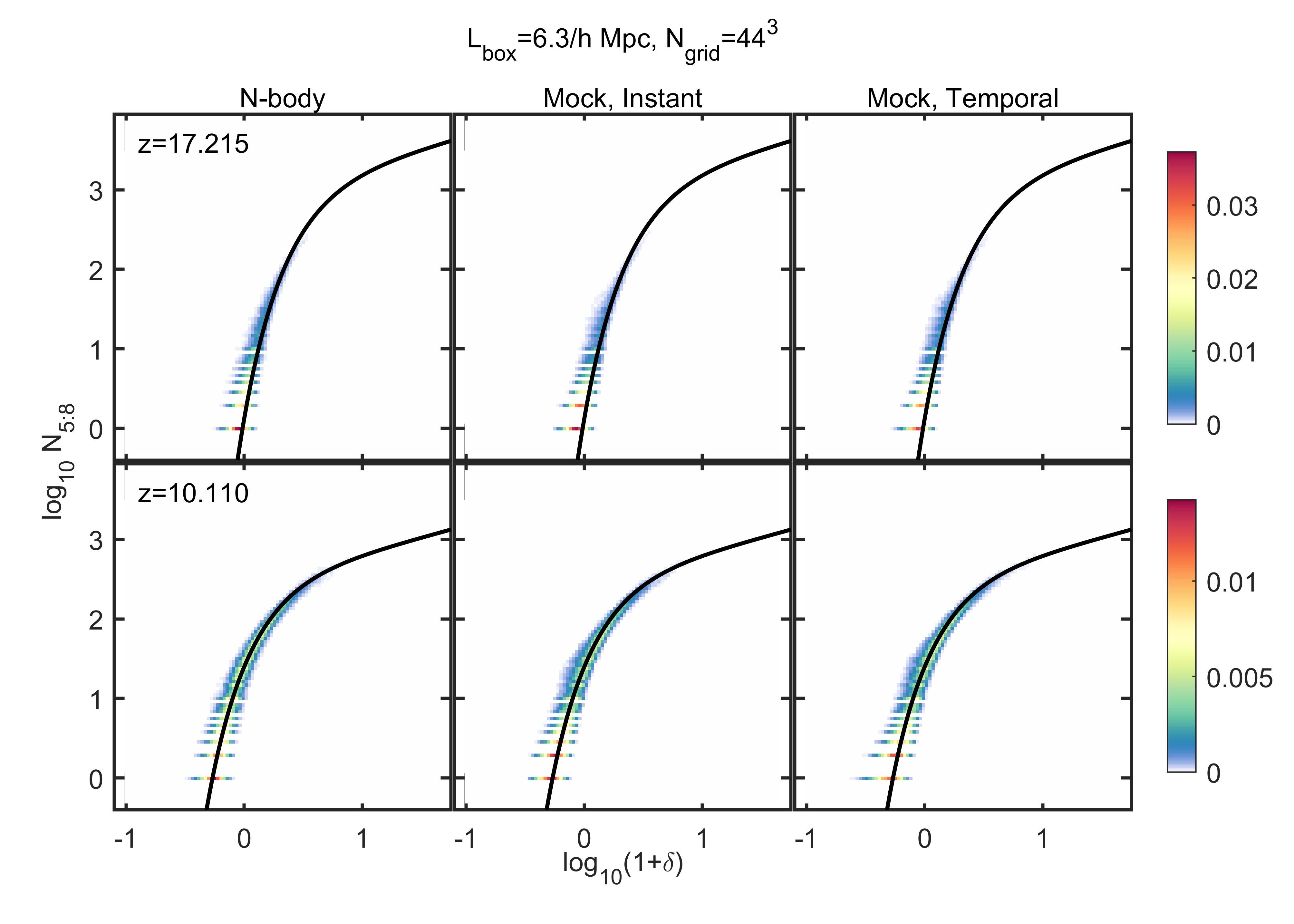}
\caption{\small\label{fig:num44} Same as Fig.~\ref{fig:num14}, but for
  44$^3$ grid (cell size $143\,\rm h^{-1}kpc$).}
\end{figure*}

\subsection{Implementing temporal halo bias}
\label{sect:temporal_mock}
We now describe our method to generate mock halo catalogues with temporal bias, based on the empirical parameters described
in Sections~\ref{sect:instant_Nbody} and \ref{sect:temporal_Nbody}. Again,
a fluctuating 3D density field should be provided at target redshifts and
the uniform Eulerian grid is used, while any quantity with a prime symbol
denotes a value related to the mock data.

\begin{enumerate}
\item Set $\Delta z$ of the density field identical to $\Delta z$ of the
  N-body simulation that were used to set parameters in
  \S~\ref{sect:temporal_Nbody}.
\item At starting redshift, which should also be identical to that of the
  N-body simulation, apply the scheme for instantaneous bias
  (\S~\ref{sect:instant_mock}) but using $f_{+}(\bar{\delta})$ instead of 
  $P_{+}(z|\delta)$ to choose the cells.
\item Presume that mock haloes were generated at $z_{\rm prev}$. Among the
  cells of given $x'=x_{\rm prev} > 0$ at $z_{\rm prev}$, randomly choose a
  fraction $P_{0}(z|x_{\rm prev},+)$ of these cells that will become empty
  at $z$, with the ``conditional de-occupation probability''
\begin{equation}\label{eqn:prob_depopulate}
P_{0}(z|x_{\rm prev},+)=\frac{N_{0}(z|x_{\rm prev}, +)}{N_{0}(z|x_{\rm prev}, +)+N_{+}(z|x_{\rm prev}, +)},
\end{equation}
and let the remaining cells host haloes again at $z$. The number of these
empty cells at $z$ given they were non-empty at $z_{\rm prev}$ chosen this way,
$N'_{0}(z|x_{\rm prev},+)$, may differ from $N_{0}(z|x_{\rm prev},+)$ in general, so similarly, $N'_{+}(z|x_{\rm prev},+)$, may also differ from $N_{+}(z|x_{\rm prev},+)$.
Let ``Group $(+|+)$'' denote the group of these cells which will still host haloes.
\item Get the total number of cells in Group $(+|+)$ from step (iii),
  $N'_{+}(z|+)=\sum_{{\rm bin\,of}\,x_{\rm prev}} N'_{+}(z|x_{\rm prev},+)$.
\item Exclude Group $(+|+)$ from the whole set of grid
  cells at $z$. Let ``Group $(0)$'' denote this group of cells. Note that Group $(0)$ also includes cells that have been de-occupied of halos at this redshift.
\item Obtain $N'_{+}(z)$ using Equation \ref{eqn:new_tot} of
  \S~\ref{sect:instant_mock}.
\item Inside Group $(0)$, try selecting $N'_{+}(z)-N'_{+}(z|+)$ cells through
  the following procedure. For each and every cell inside Group $(0)$,
  identify $f_{+}(\bar{\delta})$ as the probability for the cell to host haloes,
  where $\bar{\delta}$ is the average overdensity of the cell over $\Delta z$.
  Perform Monte Carlo sampling cell by cell to mark non-empty cells, and count
  their number $N'_{+}(A|0)$. Check if $N'_{+}(A|0)$ is within pre-set
  tolerance of the target number $N'_{+}(z)-N'_{+}(z|+)$. If not, then iterate
  the Monte Carlo procedure until convergence. Let ``Group $(+|0)$'' denote
  the resulting group of non-empty cells.
\item Finally, apply steps (iii) and (iv) in \S~ \ref{sect:instant_mock}
  on all cells of both Group $(+|+)$ and Group $(+|0)$.
\item Steps (iii) -- (viii) are done over all redshifts, taking the results
  from step (ii) as the first case of $z_{\rm prev}$.
\end{enumerate}

We note that the target number of non-empty cells at $z$ is $N'_{+}(z)$ (step
vi) is also the number of non-empty cells in the case of instantaneous
bias following Equation \ref{eqn:new_tot}. This is because the temporal
 bias prescription should of course
satisfy the instantaneous statistics at the least. We also note that the
smaller $x_{\rm prev}$ is, the larger $P_{0}(z|x_{\rm prev},+)$ becomes, implying
that cells containing a small number or mass of haloes are relatively likely to become
devoid of haloes in the near future.

\section{Results}
\label{sect:results}
 
Based on the methodology presented in \S~\ref{sect:method}, we can create
mock realizations of halo spatial distributions based on an input cosmological
density field for a given spatial resolution (cell size) and simulation
volume. Our immediate aim and first application for this method is to create
such mocks as input for large-volume radiative transfer simulations of
cosmic reionization for cases where low-mass galaxies driving reionization
cannot be resolved numerically. This is discussed in
\S~\ref{sect:eor_application} below.

However, before discussing the large-scale mocks in
\S~\ref{sect:eor_application}, we first demonstrate in \S~\ref{sect:tests} our methodology on the density fields derived from the same
high-resolution N-body simulations that provided the
fitting parameters in the first place, both with and without the temporal bias.
Since in this case we directly resolve the low-mass haloes of interest, we can
evaluate the fidelity of our mocks against the actual known halo numbers and
spatial distribution.

Furthermore, the direct comparison of the cases with and without temporal
bias helps us understand the importance of its inclusion in modelling the
stochasticity. We start sampling the data from $z=30$, but only start implementing
the stochasticity from $z \leq 25$ to ensure that there is sufficient data to
determine the parameters that are truly reflective of the distribution.\\

\begin{figure*}
\centering
\includegraphics[width=0.84\textwidth]{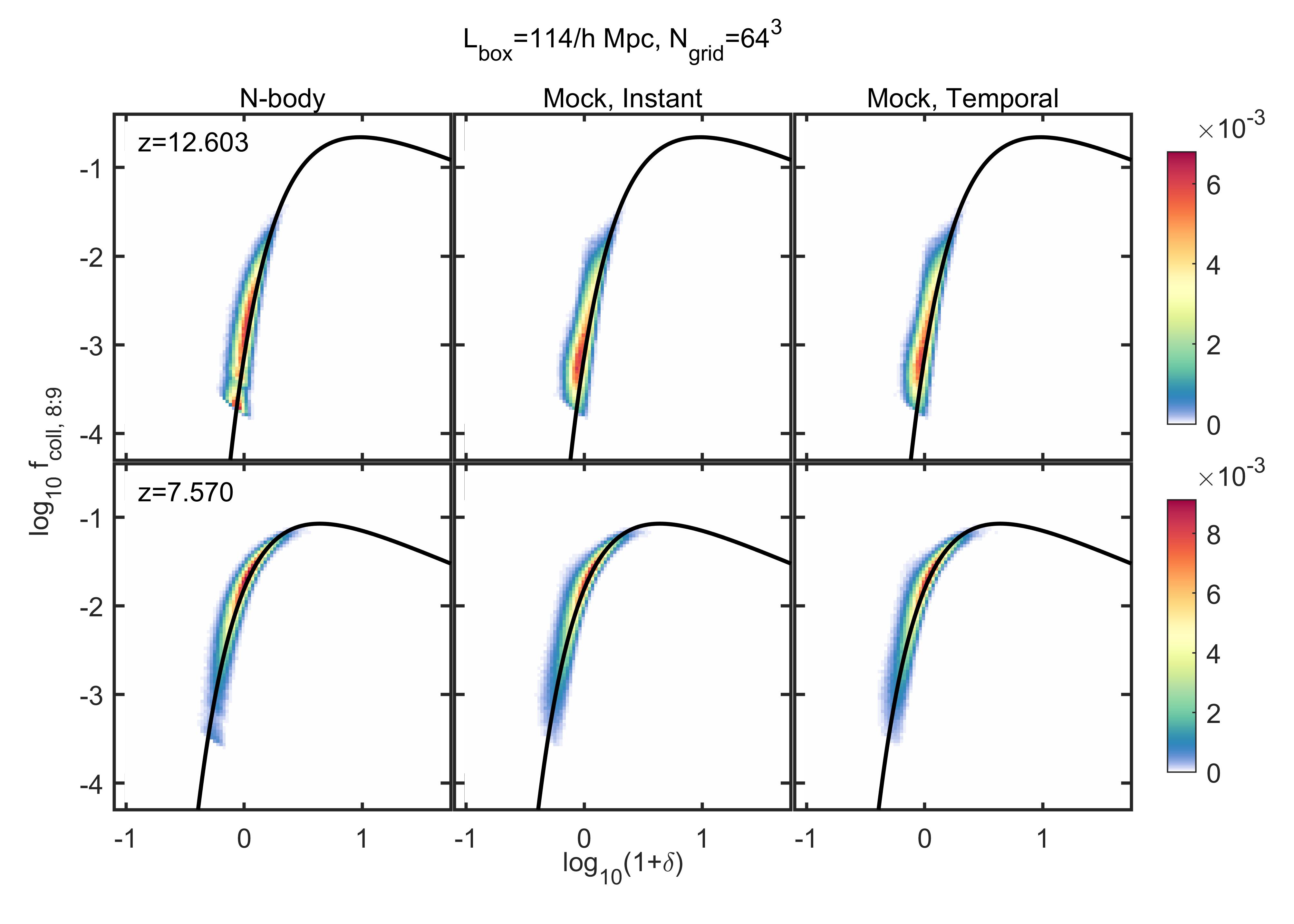}
\caption{\small\label{fig:fcoll64} LMACHs collapsed fraction, $f_{\rm coll,8:9}$,
  per cell vs cell overdensity $\delta$ at $z=12.603$ (upper panel) and
  $z=7.570$ (lower panel) for 114\,Mpc\,h$^{-1}$ box and 64$^3$ grid (cell
  size $1.781\,\rm h^{-1}Mpc$). Shown are the N-body halo data (left
  panels), the instantaneous mock haloes (middle), and
  the temporal mock haloes (right).} 
  
\centering
\includegraphics[width=0.84\textwidth]{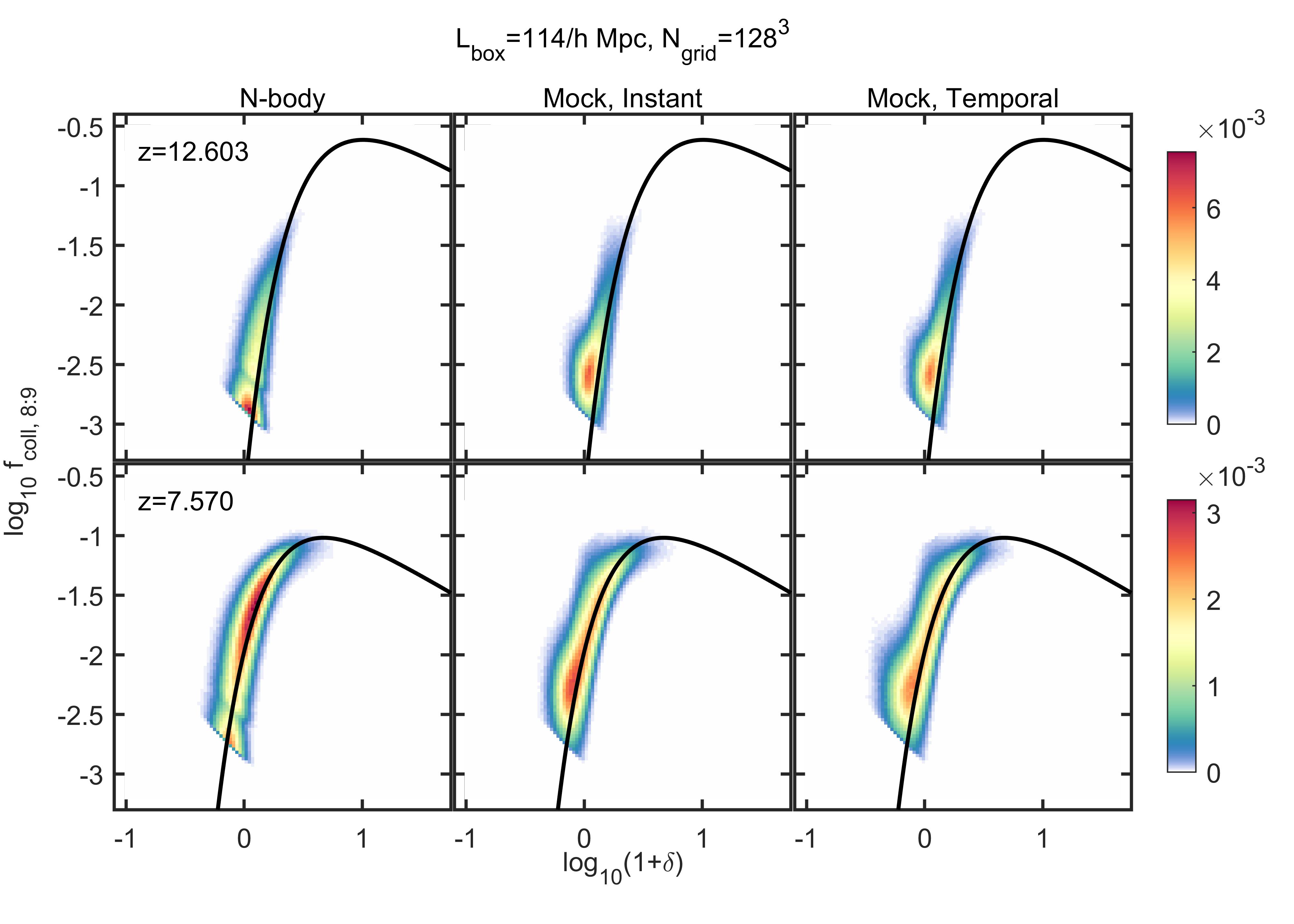}
\caption{\small\label{fig:fcoll128}Same as in Fig.~\ref{fig:fcoll64}, but for
  $128^3$ grid (cell size $0.891\,\rm Mpc\,h^{-1}$).} 
\end{figure*}

\subsection{Testing the method}
\label{sect:tests}

In Figures~\ref{fig:num14} and \ref{fig:num44}, we show the relation between
the number of MHs per cell, $N_{5:8}$, and the density of a cell, $\delta$, at $z=17.215$ (upper
panels) and $z=10.110$ (lower panels) based on data from our 6.3 Mpc $h^{-1}$ box simulation with cell sizes $450\,h^{-1}$kpc and $143\,h^{-1}$kpc, respectively. Shown are the N-body simulation data (left panels), the
instantaneous mock haloes (middle panels), and temporal mock haloes (right panels). The colours indicate the density
of haloes and the solid black lines correspond to the mean (deterministic) bias
relation based on \citet{ahn2015non}. The particular redshifts shown are chosen
as representative of the early and late stages of the epoch during which the MHs
are expected to be an important component of and a regulating factor to early
star formation \citep{2012ApJ...756L..16A}. For consistency, we will use the same colour scheme
throughout this paper to represent the different types of data.

Similarly, Figures~\ref{fig:fcoll64} and \ref{fig:fcoll128} show plots of the
LMACH collapsed fraction, $f_{\rm coll,8:9}$ vs. $\delta$ at redshifts
$z=12.603$ (upper panels) and $z=7.570$ (lower panels). Data is based on the 114
Mpc $h^{-1}$ volume with 64$^3$ and 128$^3$ grid cells per each dimension (cell sizes of 1.781\,Mpc
$h^{-1}$ and 0.891\,Mpc $h^{-1}$), respectively. We again show
the N-body simulation data (left panels), the
instantaneous mock haloes (middle panels), and temporal mock haloes (right panels). The illustrative redshifts shown roughly correspond to the beginning and the peak of
the reionization process in certain classes of EoR models.

Figures ~\ref{fig:num14} and \ref{fig:num44} show that the relation between $N_{5:8}$ and $\delta$, including stochasticity,  closely replicates that of N-body MHs once the bias is sampled from the lognormal
distribution described in the previous section. This is true for both the instantaneous and the temporal case with only minor differences. In all
cases the agreement with the mean bias trend (black curve) is excellent.

The stochasticities in the mock halo data have a good agreement with those of the N-body halo data, but with some modest difference in the low-$\delta$ regime as seen in Figures \ref{fig:fcoll64} and \ref{fig:fcoll128}.
We attribute this difference to the Poisson noise and our specific scheme of smoothing $\mu(\delta)$ in the low-$\delta$ regime, whose empirical values have a rather strong fluctuation for varying $\delta$. The Poisson noise in these cells having a small number of halos seems to bias $\mu(\delta)$ toward relatively large values, and as a result a smooth fitting function of $\mu(\delta)$ we tend to push the overall dispersion of mock data points upward in  Figures \ref{fig:fcoll64} and \ref{fig:fcoll128}. Nevertheless, these cells take only a minor fraction and thus the overall agreement seems excellent.

Credibility of the generated mock halo catalogues can be tested in terms of statistical measures, and we investigate (1) overall normalisations (either number of haloes or total collapsed fraction) for the deterministic bias, (2) the temporal correlation coefficient for the cell-wise evolution, and (3) the power spectrum for the spatial clustering of haloes.

\begin{figure*}
\centering
\includegraphics[width=0.85\textwidth]{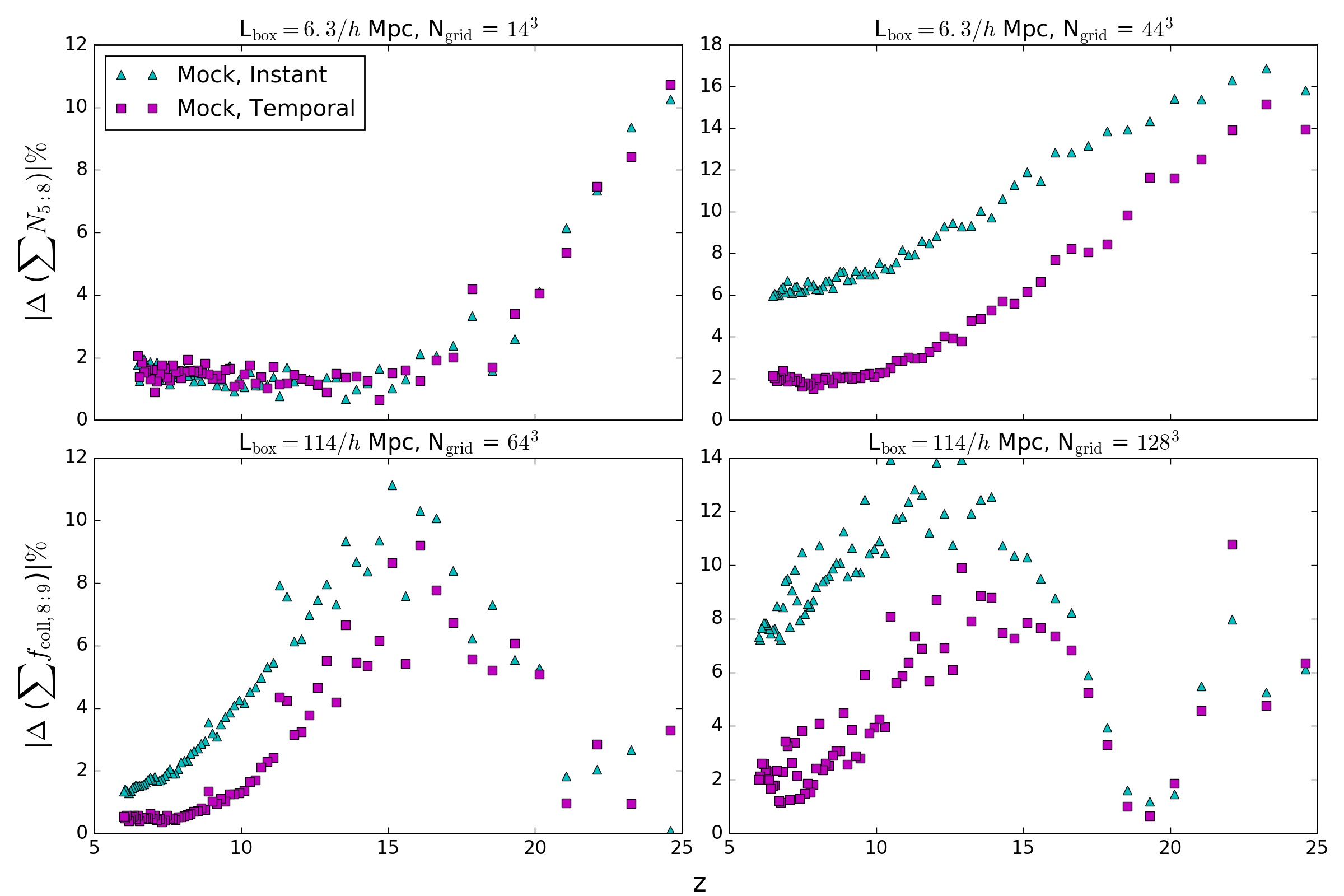}
\caption{\small\label{fig:total_scatter} Evolution of the absolute value of the
  percentage difference between the N-body data and the total number of MHs, $N_{5:8}$, (upper panels) and
  the LMACH collapsed fraction, $f_{\rm coll,8:9}$, summed over the grid (lower panels)
  vs. redshift, $z$, for the 6.3 Mpc$\,h^{-1}$ and 114 Mpc$\,h^{-1}$ box respectively.
  Plotted are the difference between the N-body data and the instantaneous mock haloes (triangles) and the temporal mock haloes (squares) for
  the 14$^3$, 44$^3$, 128$^3$ and 64$^3$ grid, as labelled. 
}
\end{figure*}

\begin{figure*}
\centering
\includegraphics[width=0.85\textwidth]{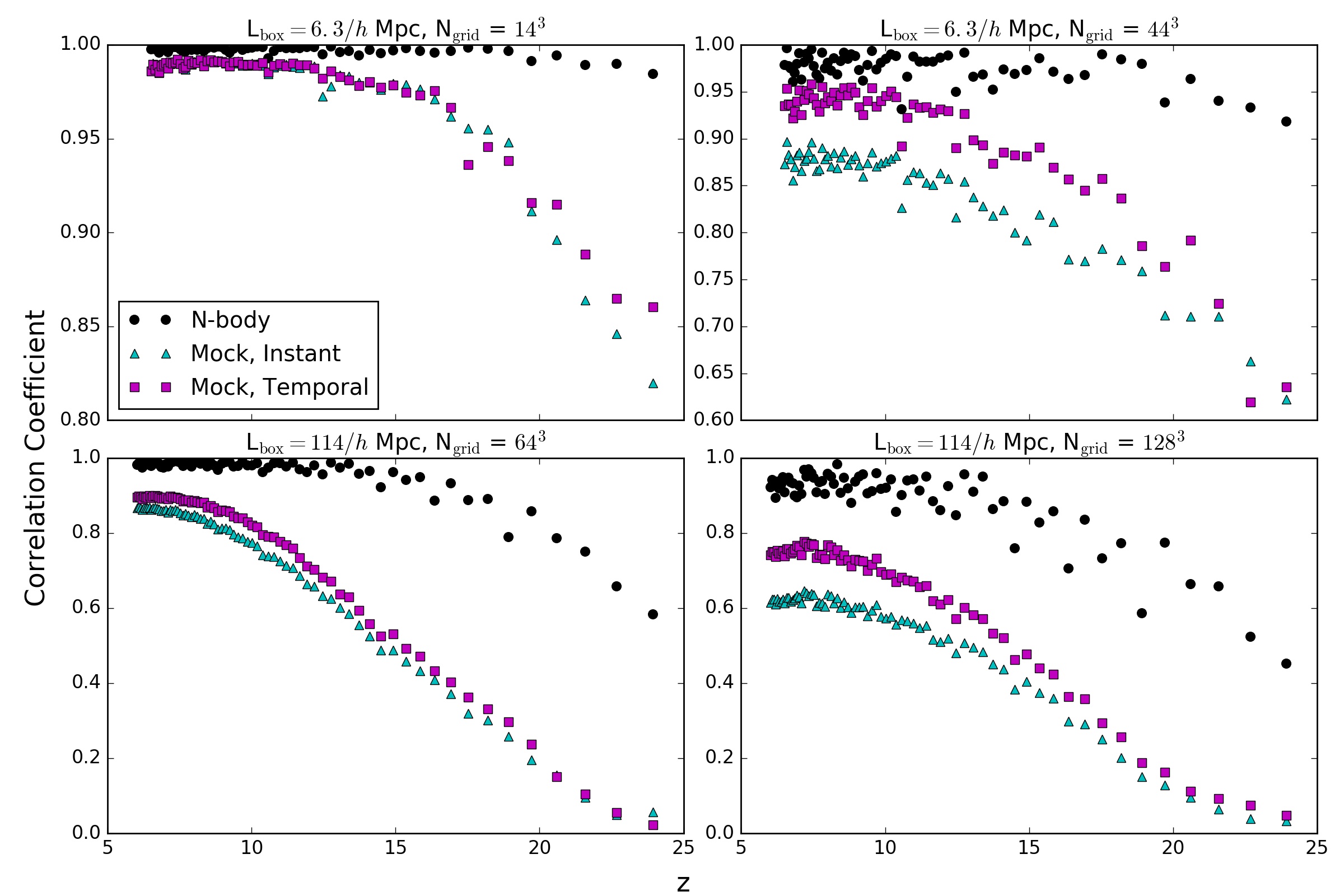}
\caption{\small\label{fig:corrcoeff}
Pearson correlation coefficient between consecutive time slices vs redshift $z$
for: (top) 6.3 Mpc$\,h^{-1}$ box with (left) grids of 14$^3$, and (right) 44$^3$,
for the simulation data, and realisations of instantaneous mock and temporal mock haloes as
labelled; (bottom) same as the top panels, but for 114 Mpc$\,h^{-1}$ box with (left)
grids of 64$^3$ and (right) 128$^3$.
}
\end{figure*}
We show the overall normalisations and their evolution with
redshift in Figure ~\ref{fig:total_scatter}. In all cases,
difference between the mock data and the N-body halo data are modest, within $\sim 15\%$, and often substantially better,
particularly at lower redshifts. The mocks with temporal bias consistently
match the simulations more closely than those with instantaneous bias, within $\sim 2\%$ below $z=10$ and
otherwise within 5\% or less except for the highest redshifts, 
where the rarity of haloes make the correlation with the underlying density field weak. The mocks follow the
N-body halo data better for the lower-resolution grids for each volume, and
the temporal bias yields little difference for the number of MHs in that
case. The instantaneous bias is considerably worse than the temporal bias in normalisations by up to a factor of 2-4, especially for higher
grid resolutions.

Next we compare the Pearson correlation coefficient between consecutive
time-slices of the simulation data and the respective generated stochastic
realisations. The correlation coefficient is as usual defined as
\[
r_{X,Y}=\frac{cov(X,Y)}{\sigma_X\sigma_Y}
\]
where $X$ and $Y$ are the fields being correlated, $cov$ is the covariance
matrix, and $\sigma_{X,Y}$ are the standard deviations of $X$ and $Y$,
respectively. The value of $r_{X,Y}$ is $1$ if $X$ and $Y$ are completely
correlated, $0$ if they are uncorrelated and $-1$ if they are completely
anti-correlated. Results are shown in Fig.~\ref{fig:corrcoeff}. For the
simulation data, the late-time correlation is fairly tight in all cases.
At early times the correlation coefficient is considerably lower than unity
due to the increasing rarity of haloes, and the related significant Poisson
noise. At that epoch, the halo population, which form stochastically in
high-density regions, grows exponentially, thereby resulting in the lower
correlation with the previous time-slice. This effect is larger for more
massive haloes, as could be expected, as they are more strongly biased.

Randomly-sampled mocks without the temporal correlation bias significantly
underestimate the correlations compared to the simulation data, and yield
almost no correlation for LMACHs at high redshift (independent of the grid
resolution). As could be expected, in all cases the agreement is significantly
improved by including the temporal bias, although the correlation remains
somewhat lower than the simulated one. This agreement could potentially be
improved further by modifying our model to take into account the tight
empirical time correlation found in the simulation data, but at the expense of a more complex model. 

\begin{figure*}
\centering
\includegraphics[width=0.85\textwidth]{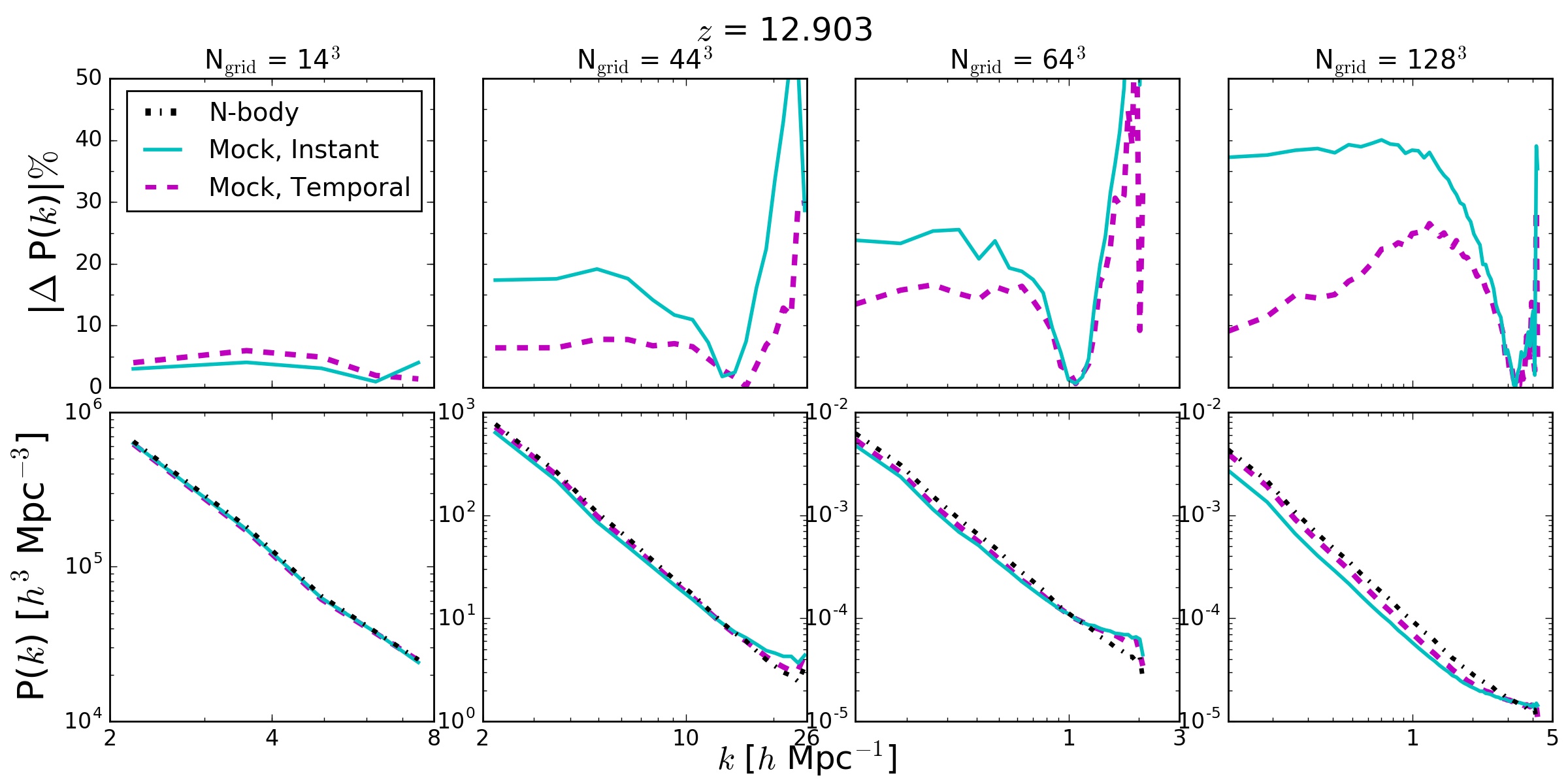}
\caption{\small\label{fig:PS-1D-2} The lower panels show the $P(k)$ and the
  upper panels show the percentage of absolute difference of the $P(k)$ between the N-body data and the mock realisations for the respective grids at $z= 12.903$. The black dash-dot
  line corresponds to the N-body data, the cyan solid line
  corresponds to the generated instantaneous mock haloes and the
  magenta dash line corresponds to the generated temporal mock haloes. The percentage in the upper panel is capped at $50\%$.
  }
\end{figure*}

\begin{figure*}
\centering
\includegraphics[width=0.85\textwidth]{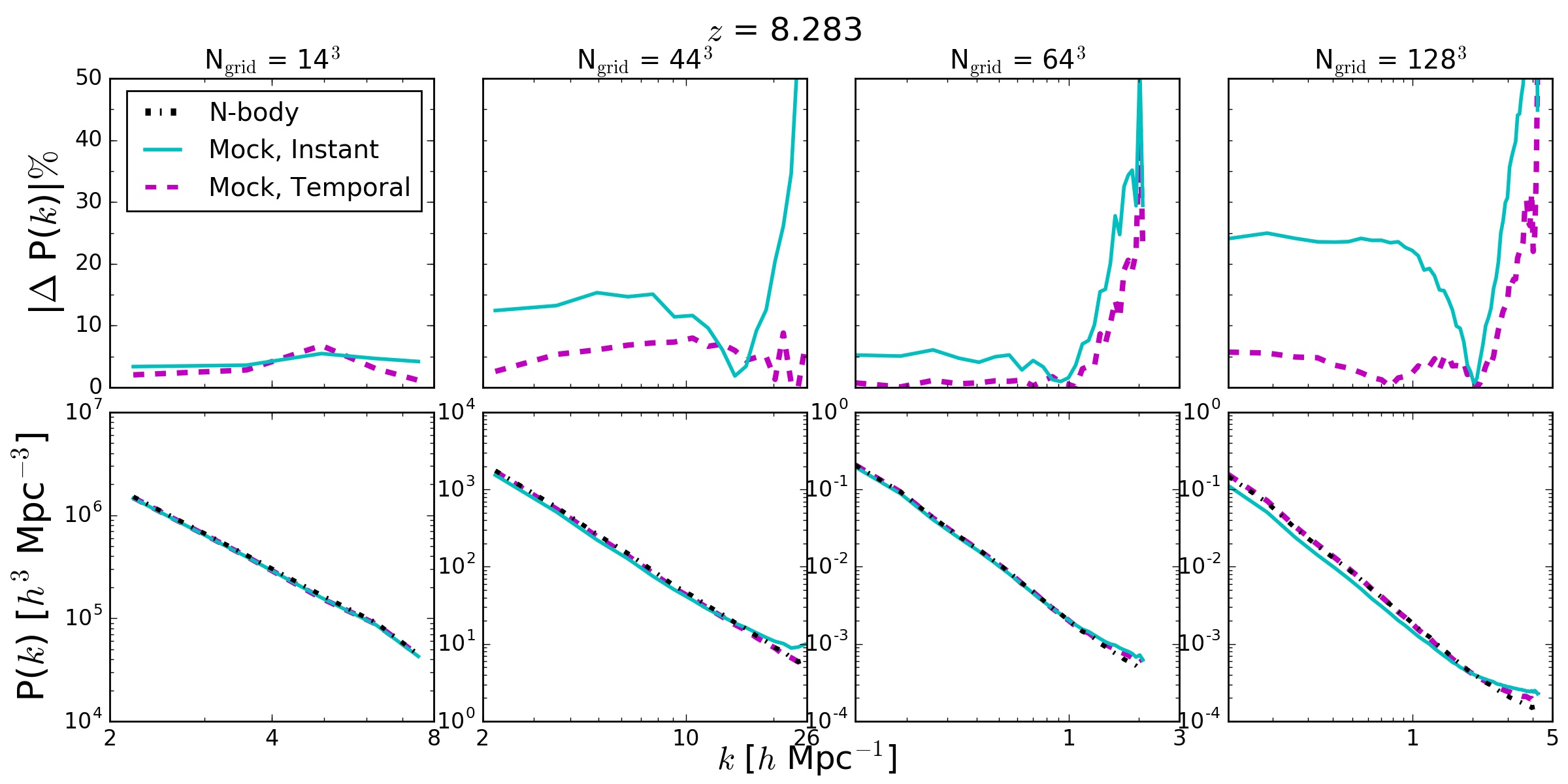}
\caption{\small\label{fig:PS-1D-1} Same as Figure \ref{fig:PS-1D-2} but for
  $z=8.283$. 
  }
\end{figure*}

Finally, we consider the halo spatial clustering in each case, as represented
by the spherically-averaged power spectrum $P(k)$ which is shown in
Figures~\ref{fig:PS-1D-2} (for $z=12.903$) and \ref{fig:PS-1D-1} ($z=8.283$). In these figures, the lower panels show $P(k)$, and upper panels show the
percentage differences between the mocks and the N-body halo data, for all
cases as labelled.
The halo clustering of the mock realizations with temporal bias generally matches
the simulated one very well, within a few to 10 per cent. The only
exceptions are at small scales ($k\gtrsim 10 h {\rm Mpc}^{-1}$) and at very high
redshift ($z>20$ for MHs and $z>15$ for LMACHs), due to the aforementioned
rarity of haloes at high redshift and the consequent Poisson
noise, in which cases the discrepancy can reach $\sim 50\%$. The agreement
is considerably worse in general when the temporal bias is not included,
with typical differences of $\sim$20-50\%. The only exception is the MH numbers in the
low-resolution grid, where the agreement becomes better, indicating
that the halo clustering is insensitive to the temporal bias in this limited case. 

\subsection{Sub-grid Modelling of Temporal Stochasticity in Multi-Scale
  reionization}
\label{sect:eor_application}

\begin{figure*}
\centering
\includegraphics[width=0.84\textwidth]{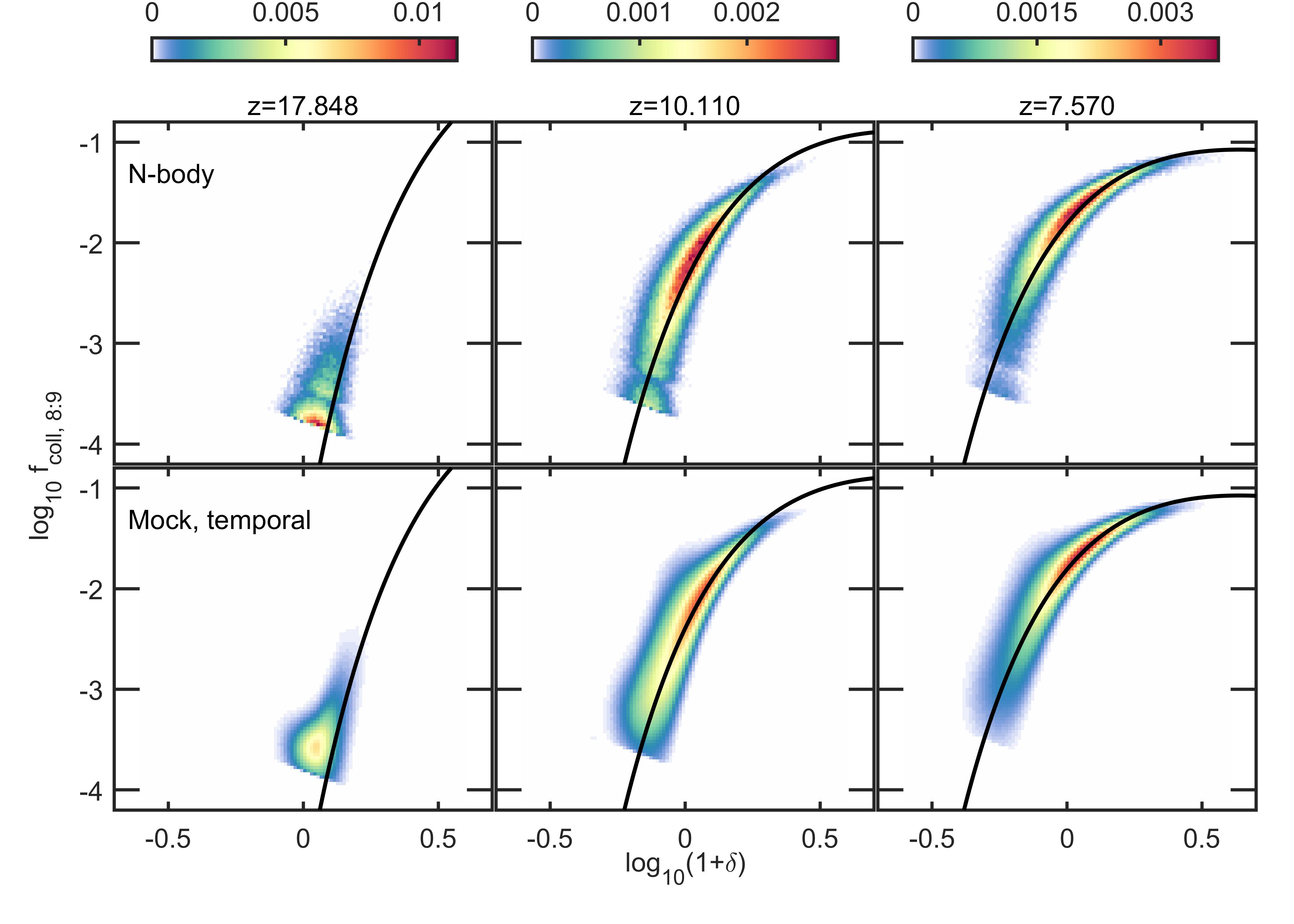}
\caption{\small\label{fig:realization500}The figure shows the
  LMACH collapsed fraction $f_{\rm coll,8:9}$ per cell with respect to $\delta$ at
  $z=17.848$, $10.110$  and $7.570$ (left to right). The top
  panels correspond to the actual N-body simulation data from our
  $114\,h^{-1}$~Mpc box on a 64$^3$ grid, while the bottom
  panels show the generated mock haloes with temporal bias for
  $500\,h^{-1}$~Mpc box on a 300$^3$ grid. The color map shows
  the normalized density of data points in discretised bins of
  $\log_{10}(1+\delta)$ and $\log_{10}f_{\rm coll,\,8:9}$.
}
\end{figure*}

We implement our fiducial method, namely assigning stochasticity sampled
from a log-normal distribution with temporal correlations
(\S~\ref{sect:temporal_mock}) in generating mock halo catalogues on a
$300^3$-grid density field in a large volume of 500 Mpc$\,h^{-1}$ per side
simulated with the CubeP$^3$M code \citep{2013MNRAS.436..540H}. 
We have previously shown that a 500 Mpc$\,h^{-1}$ volume is sufficiently large
to obtain very reliable statistics in various  physical properties of EoR
\citep{2014MNRAS.439..725I}, and a 300$^3$-grid is optimal for simulating
patchy reionization with reasonable computational resources. The size of
the grid cell is also chosen to match the cell size of the $64^3$-grid
in 114 Mpc$\,h^{-1}$-box we used above for the LMACHs collapsed fraction. This
approach enables populating such a large volume, which is usually limited
in realizing small-mass haloes due to numerical resolution limit, with
LMACHs very reliably as shown in the previous sections.

In Figure~\ref{fig:realization500} we show the modelled stochasticity with
temporal correlations of the $f_{\rm coll;8:9}$ over a range of redshifts,
representative of the early, middle and late phases of reionization,
respectively. The overall characteristics of the mock data are in good
agreement with those of the N-body data, as intended. While not clearly
shown in Figure~\ref{fig:realization500}, the large volume contains cells
with $\delta$'s in the more extended tail ends of the PDF than the small box.
We note that it is again difficult to simulate the apparent dip (dubbed as
``indentation'' in \S~\ref{sect:tests}) shown in the N-body data with this
prescription.

\begin{figure*}
\centering
\includegraphics[width=0.85\textwidth]{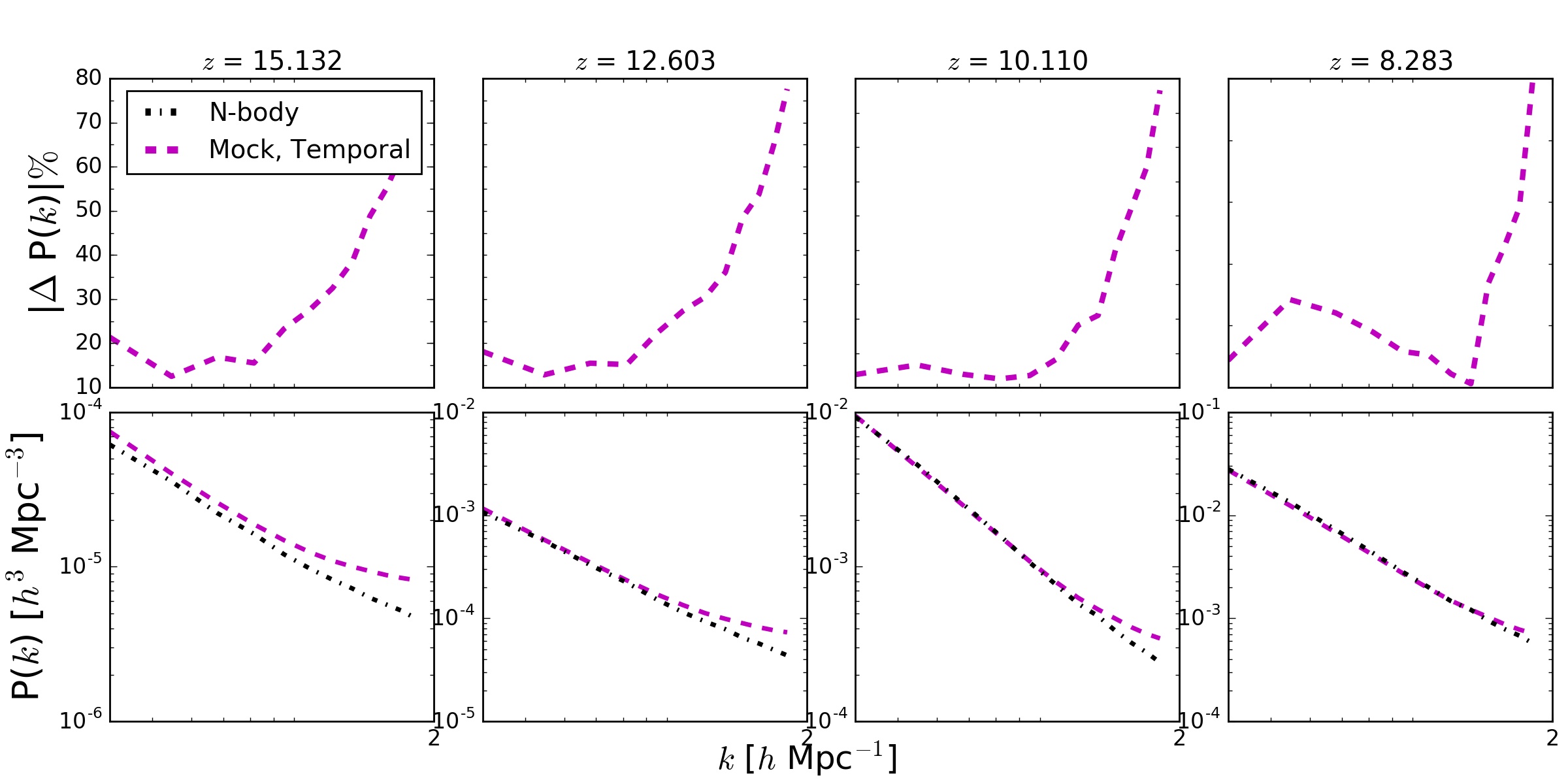}
\caption{\small\label{fig:PS-1D-realized} The lower panels show the $P(k)$
  and the upper panels show the percentage of absolute difference of the
  $P(k)$ of $f_{\rm coll;8:9}$ at the indicated redshift between the N-body (114 Mpc$\,h^{-1}$,
  $64^3$ grid, black dotted in lower panel) and mock data (500 Mpc$\,h^{-1}$, $300^3$ grid; magenta dashed).}
\end{figure*}

We compare the power spectra generated from the mock vs. N-body data
(Figure~\ref{fig:PS-1D-realized}) to check the reliability of our
approach for future reionization simulations. For the range of redshifts
shown and wavenumbers $k\lesssim 1\, h^{-1} $ Mpc, the error is
no larger than $\sim 10-30\% $. At the same wavenumber range, $P(k)$ of
the mock data has a trend to be larger at high $z$ but gradually
shifting to be smaller at low $z$ than that of the N-body data. 
Disagreement always exists for $k$'s around the Nyquist value, whose impact
on reionization requires further investigation. However, we expect that the
influence of this will be minor if one is only interested in relatively
large-scale ($k\lesssim 1\, h^{-1} {\rm Mpc}$) phenomena.

\section{Summary and Discussion}
\label{sect:summary}
In this work we present a novel scheme developed for creating
realistic sub-resolution low-mass halo populations in large-scale N-body
simulation volumes. The mock haloes produced by our method reproduce
the correct average local numbers or collapsed fraction of the unresolved
haloes very well, down to the smallest halo masses, as well as their spatial clustering
and local population evolution. This new halo bias prescription can be used
as a sub-grid model for studies of structure formation for a range of
applications that would otherwise be limited in dynamic range, thereby
enabling studies of the small-scale structures and their impact within large,
cosmological volumes.

We achieve this by including not only the deterministic bias as a function of
the local density environment \citep{ahn2015non}, but also an additional
stochastic bias, which accounts for its other dependencies, which yields a more
natural representation of the universe. Using very high-resolution N-body data,
we have compared two distinct methods for realising this stochastic bias: (1)
stochasticity reconstruction based on the current cell overdensity $\delta$
only and (2) one based on both $\delta$ and the past history of the halo
population in that locality. We found the latter method, which we dubbed
``temporal bias'', is superior to the former. Including the temporal bias yields
mock halo catalogues with significantly better statistical properties than
without it in terms of the local halo population evolution as well as the
power spectrum $P(k)$ of the 3D halo-number field, both compared to the
high-resolution N-body data where the relevant halo range is resolved.

There are a number of possible applications for our method. A direct
application of this temporal bias is in simulations of cosmic reionization.
They require very large dynamic range, since the H~II region expansion is
driven by low-mass galaxies, while proper statistics of the patchiness
demands very large volumes to be followed \citep{2014MNRAS.439..725I}.
Importantly, in this case, the evolution is cumulative and depends on the
temporal evolution of structure formation. This is in contrast to e.g.
cosmology studies with galaxy surveys, where only an instantaneous halo
bias is necessary at any observed redshift. In the future, we will use 
these results to perform very large-box simulations of cosmic reionization
with minimum halo mass of $10^5-10^{8} \msun$ and $500\,h^{-1} {\rm Mpc}$
volume, which requires applying our scheme to model the haloes with masses
below $10^9\,M_\odot$ and thus to overcome the numerical resolution limit.
The mock halo catalogue of LMACHs we generated for this volume, presented
in \S~\ref{sect:eor_application} is promising, in a reasonable agreement
in terms of its statistical properties, especially $P(k)$ of the 3D
collapsed fraction field, compared to N-body halo data from a smaller
volume ($114\,h^{-1} {\rm  Mpc}$), where LMACHs haloes are directly resolved.

This study shows that the temporal correlation in stochasticity plays an
important role in shaping the statistical properties of cosmological
haloes. This is proven by the fact that the temporal bias (method 2 above)
generates halo catalogues in much better agreement with the N-body halo
data than method (1) in terms of $P(k)$ and cross-correlations. This also
implies that even when one is to generate halo catalogues for the study of
galaxy surveys, which seemingly requires an instantaneous halo bias, this
temporal bias scheme can work as a very reliable solution. Further study
along this line is warranted.

There are some caveats and room for improvement in our methodology.
In our current approach, the temporal bias is based on the complete
history of a given Eulerian cell, while in reality the stochasticity
should only depend on the past history. Furthermore, advection of
matter to and from neighbouring cells means that the Eulerian cell
density does not contain the full information on the matter field
evolution. For a better temporal bias prescription, one may instead
adopt a scheme that takes a limited lookback-time history of an
Eulerian cell to mitigate these two problems. This requires further
investigation, which we will address in the near future. While our
approach and results are general, the specific paramaterisation of
the temporal bias is based on empirical fits based on a specific
structure-formation simulation. Therefore, if one were to apply this
to a universe described by a different set of cosmological parameters,
our approach would require a new small-box, high-resolution simulation
resolving haloes of our interest. It would be preferable to find a more
analytical scheme that allows {\em deterministic temporal stochasticity}
in halo bias that could be easily re-calculated.

Regardless of this, our scheme provides significant improvement over
existing methods, yielding a more accurate mock halo catalogues at
high redshift, which will be very helpful in improved descriptions
of the cosmic reionization process and interpretation of high-redshift
observations. Using this approach will yield better answers on how much
impact low-mass haloes, which have usually been neglected or treated with
crude approximations, have on structure formation at high redshift and the
history of cosmic reionization.

\section*{Acknowledgements}
AN is supported by the Australian Research Council Centre of 
Excellence for All Sky Astrophysics in 3 Dimensions (ASTRO 3D), 
through project number CE170100013. KA was supported by 
NRF-2016R1D1A1B04935414. This work was supported by the
Science and Technology Facilities Council [grant number ST/I000976/1] and the
Southeast Physics Network (SEPNet). We acknowledge that the results in this
paper have been achieved using the PRACE Research Infrastructure resource
Marenostrum based in the Barcelona Supercomputing Center, Spain. Time on this
resource was awarded by PRACE under PRACE4LOFAR grants 2012061089 and
2014102339 as well as under the Multi-scale reionization grants 2014102281
and 2015122822. The authors gratefully acknowledge the Gauss Centre for
Supercomputing e.V. (www.gauss-centre.eu) for funding this project by providing
computing time through the John von Neumann Institute for Computing (NIC) on
the GCS Supercomputer JUWELS at J\"ulich Supercomputing Centre (JSC). Some of
the numerical computations were done on the Apollo cluster at The University
of Sussex.

\bibliographystyle{mn2e} \bibliography{refs1.bib}

\label{lastpage}
\end{document}